\documentclass[pre,aps,twocolumn,showkeys,amssymb]{revtex4}
\usepackage{epsfig}
\usepackage{dsfont} 

\begin{document}

\title{A novel stochastic Hebb-like learning rule for neural networks}
\author{Frank Emmert-Streib}
\email{femmert@physik.uni-bremen.de}
\affiliation{Institut f\"ur Theoretische Physik, Universit\"at Bremen, Otto-Hahn-Allee, 28334 Bremen, Germany}

\date{\today}

\begin{abstract}
 
We present a novel stochastic Hebb-like learning rule for neural networks. This learning rule is stochastic with respect to the selection of the time points when a synaptic modification is induced by pre- and postsynaptic activation. Moreover, the learning rule does not only affect the synapse between pre- and postsynaptic neuron which is called homosynaptic plasticity but also on further remote synapses of the pre- and postsynaptic neuron. This form of plasticity has recently come into the light of interest of experimental investigations and is called heterosynaptic plasticity. Our learning rule gives a qualitative explanation of this kind of synaptic modification.

\end{abstract}

\maketitle

\section{Introduction}

What are the mechanisms that modulate learning on a neuronal level in animals or humans? This question is up to now under debate, but the imagination one has for a biological learning rule is that the synaptic weights are changed according to a local rule. In the context of neural networks local means that only the adjacent neurons of a synapse contribute to changes of the synaptic weight. Such a mechanism with respect to synaptic strengthening was proposed by Donald Hebb \cite{h1949} in 1949 and experimentally found by T. Bliss and T. Lomo \cite{blisslomo_1973}. In a biological terminus Hebbian learning is called {\it{long-term potentiation}} (LTP).

Experimentally as well as theoretically there is a great body of investigations aiming to formulate precise conditions under which learning in neural networks takes place. E.g. the influence of the precise timing of pre- and postsynaptic neuron firing \cite{markramluebke_1997,kempter_1999} or the duration of a synaptic change (for a review see \cite{koch_1999}) termed {\it{short}} or {\it{long-term plasticity}} have been studied extensively. All of these analyses share the locality condition proposed by Hebb \cite{h1949}. 

But there are also experimental findings which extend the traditional view of synaptic plasticity in three important points. First Frey and Morris \cite{freymorris_1997} found in the hippocampus of rats in vivo that there is a {\it{synaptic tagging}} mechanism. This mechanism tagges synapses which were repeatly involved in information processing within a certain time window of up to 1.5h. If one of these synapses is restimulated within this time interval then LTP is induced. Thus they concluded that there is a form of memory for past synaptic activity which leads to a kind of summing up past activities to induce LTP. The second result is from Otmakhova and Lisman \cite{otmakhova_1998} who found an influence of a global dopamin signal on LTP and LTD in CA1 in the hippocampus. A third extention consists in results about heterosynaptic plasticity. In this form of learning not only the synapse between active pre- and postsynaptic neuron is changed but also further remote synapses of these neurons \cite{bipoo_2001}. Heterosynaptic plasticity is observed for LTP as well as LTD.

We emphasize that none of these additional findings exclude the classical locality condition by Hebb but involve further contributions to specify learning more precisely.

Based on the exciting results of Frey and Morris \cite{freymorris_1997} and Otmakhova and Lisman \cite{otmakhova_1998} there are theoretical investigations on learning dynamics in neural networks which interweave the locality condition of Hebb with a {\it{synaptic tagging}} mechanism and a global control signal. Chialvo and Bak \cite{cb1999,bc2001} suggested a learning rule which assigns each synapse a boolean scalar valued variable indicating if the synapse were involved in the last information processing or not. This mimics the tagging mechanism. Additionally they assigned to the global control signal the role of an external reinforcement signal $r$ which forms a kind of feedback for the network performance. The reinforcement signal can also take only two different values whereas $r=1$ corresponds to a right and $r=-1$ to a wrong output of the network. Synaptic update was only allowed if the synapse was activated during the last signal processing and the reinforcement signal $r$ signaled a failure due to a wrong output of the network. An extention of the learning rule of Bak and Chialvo was presented by Klemm, Bornholdt and Schuster \cite{kbs2000}. They allowed a synaptic memory $c$ for each synapse with $\Theta+2\in \mathds{N}$ discrete values. The dynamics of the synaptic counter $c$ is in each time step given by $c_{t+1}=c_t-r_t$ for active synapses which is restricted from below to $0$. If $c_t-r_t>\Theta$ occurs because the output of the network was wrong and hence a reinforcement signal $r_t=-1$ was fed back into the network then the corresponding synapse will be depressed by a fixed amount $\delta$ and the corresponding synaptic counter is set to  $c_t-r_t=\Theta$. 

In this paper we present a novel Hebb-like learning rule which has a memory for the past failures similar to \cite{cb1999,kbs2000,bc2001}. However, in contrast to these works we do not use synaptic but neuron counters. Due to the use of neuron counters which can be seen as approximation for the synaptic counters we are lead to a stochastic update condition instead of a deterministic one for active synapses. The obtained stochastic learning rule whose character is still local can be interpreted biologically and corresponds in a qualitative way to heterosynaptic plasticity \cite{bipoo_2001}.

The paper is organized as follows. In section  \ref{model_xor} we describe a model with which we investigate our stochastic learning rule. The learning rule itself is motivated and defined in section \ref{defslr}. The result section \ref{results} is subdivided in three parts. Because a learning rule of a neural network is only one part of the entire system we investigate the interplay between our stochastic learning rule and three different network dynamics and hence their influence on the convergence behavior of the neural network. We compare a winner-take-all \ref{wta}, a softmax \ref{softmax} and a noisy winner-take-all mechanism \ref{nwta} which are all different forms of lateral inhibition. In section \ref{nwta} we investigate additionally the influence of a variable size of a synaptic change $\delta$. The results are discussed and compared with \cite{kbs2000}. A biological interpretation of our stochastic Hebb-like learning rule with respect to heterosynaptic plasticity is given in \ref{biointerpretation}. The paper ends in section \ref{conclusions} with a summary and conclusions.


\section{The model}\label{model_xor}

To investigate the learning dynamics of a neural network one has to define every item in table \ref{gen_sys}.
\begin{table}[h!]
\begin{tabular}{cl}
1.& topology of the neural network\\
2.& neuron model\\
3.& network dynamics\\
4.& learning rule\\
5.& environment\\
6.& interaction of the TBM with the environment\\
\end{tabular}
\caption{\label{gen_sys}Characterization of the entire system}
\end{table}

With Toy-Brain-Model (TBM) we summarize the points 1.) to 4.) in table \ref{gen_sys}. The concrete definitions for each part are as follows. 1.) Topology of the neural network: We choose a feedforward network with three layers. The layers consist of $I$ input-, $H$ hidden- and $O$ output neurons. The neurons of adjacent layers are all to all connected with synapses $w_{ij}\in\mathds{R^+}$. 2.) Neuron model: The neurons are binary $x_i\in\{0,1\}$ with $i\in\{1,\ldots,I+H+O\}$. As network dynamics we use three different types to investigate the interplay with our learning rule. We use a winner-take-all, a softmax \cite{kbs2000} and a noisy winner-take-all mechanism \cite{cb1999,bc2001}. In all three cases only one neuron is chosen to be active in the hidden and output layer according to the network dynamics. This corresponds to a low activity limit which was in \cite{cb1999} called {\em{extremal dynamics}}. 3.a) Network dynamics (winner-take-all): The inner fields of the neurons are calculated by 
\begin{eqnarray}
h_j=\sum^\mathrm{all}_i w_{ji}x_i.\label{innerfield}
\end{eqnarray}
Here {\em{all}} means all neurons of the preceding layer. The active neuron in each layer is simply the one with the highest inner field
\begin{eqnarray}
i_\mathrm{max}&=&\mathop{\mathrm{argmax}}_{i}(h_i)
\end{eqnarray}
which is set to $x_{i_{\mathrm{max}}}=1$. All other neurons are set to zero. The winner-take-all mechanism is a purely deterministic selection mechanism and uniquely determined by the inner fields of the neurons.

3.b) Network dynamics (softmax): The inner fields of the neurons are calculated by equation \ref{innerfield} but the activity $x_j$ of the neurons is now obtained by choosing one neuron from the probability distribution 
\begin{eqnarray}
p_j=Z^{-1}\exp({\beta}h_j) \label{swta}\\
Z=\sum_j \exp({\beta}h_j)   \label{swtan}
\end{eqnarray}
The activity of the chosen neuron is set to one and the other neurons are set to zero. The temperature-like parameter $\beta^{-1}(t)=\beta^{-1}\in\mathds{R^+}$ is held constant. One can regulate by $\beta$ the stochastic character of \ref{swta},\ref{swtan} because for $\beta=0$ one obtains $p_j=\frac{1}{H}$ for the hidden and $p_j=\frac{1}{O}$ for the output layer for all $j$ in the respective layer which corresponds to equal distributions. Whereas $\beta\rightarrow\infty$ results in a deterministic selection of the neuron with the highest inner field in each layer which is equivalent to the winner-take-all network dynamics.

3.c) Network dynamics (noisy winner-take-all): The inner fields  of the neurons are again calculated by equation \ref{innerfield}. In this case, the active neuron of each layer $x_{i_{\mathrm{max}}}=1$ is the one with the highest value after the addition of noise.
\begin{eqnarray}
\overline{h}_i&=&h_i+\eta_i  \label{nwta_c4}\\
i_\mathrm{max}&=&\mathop{\mathrm{argmax}}_{i}(\overline{h}_i) \label{nwtas_c4}
\end{eqnarray}
The noise $\eta_i$ is uniformly drawn out of $[0,\eta]$. Again one can by $\eta$ regulate the stochastic character of the selection mechanism and obtains for $\eta\rightarrow0$ the deterministic winner-take-all mechanism.

The definition of the learning rule is postponed to subsection \ref{defslr} because this is the central point of this paper. 

5.) Environment: We choose as problem to be learned by the network the exclusive-or (XOR) mapping shown in table \ref{xor_map}. 

\begin{table}[h!]
\begin{tabular}{lll|ll}
$x_3$&$x_2$&$x_1$&$x_8$&$x_7$\\ \hline
0&0&1&1&0\\
0&1&1&0&1\\
1&0&1&0&1\\
1&1&1&1&0\\
\end{tabular}\\
\caption{\label{xor_map}Exclusive-or (XOR) mapping}
\end{table}
Here $x_1$ is a bias introduced to exclude the case of zero activity in the input and hence in all subsequent layers. We have chosen the exclusive-or (XOR) mapping for two reasons. First, the problem is not linear separable, it can not be learned by a single perceptron \cite{minskypapert_1969} but only by a multilayer network. However, up to the discovery of the back-propagation algorithm of Rumelhardt, Hinton and Williams \cite{rumelharthinton1_1986} in the 80's there was no systematic method known to adjust the synaptic weights of the neural network. Still, the problem with the back-propagation algorithm is that it is not biological plausible because it requires a back propagation of an error in the network which can not be known \cite{crick_1989}. For this reason learning by back-propagation is classified as supervised learning \cite{hertzkrogh_1991}. Second, the biologically plausible learning rules proposed by \cite{kbs2000} and \cite{bc2001} demonstrate that they are able to cope with the exclusive-or (XOR) problem.

We call the exclusive-or (XOR) mapping the environment of the neural network in order to keep in mind that living beings with brains are always situated in an environment in which they live. The abstract environment in which our TBM lives is the exclusive-or (XOR) problem. In the context of Artifical Intelligence or Computational Neuroscience this is called embodiment \cite{brooks_1991,varela_1991}. 

6.) Interaction of the TBM with the environment: Every input pattern in table \ref{xor_map} is presented with equal probability and independent of preceding patterns.

All the above defined points form the framework of this paper. We will now turn to the motivation and definition of the learning rule.



\subsection{Definition of the stochastic Hebb-like learning rule}\label{defslr}

The question we would like to answer with respect to the proposed learning rules of \cite{cb1999,bc2001,kbs2000} is: Can the idea of a synaptic counter be simplified? Let us consider the consequences of the occurrence of the synaptic memory in biological living beings. According to \cite{churchland_1992} there are $10^{15}$ synapses in the human and $10^{13}$ synapses in the rat's brain but only $10^{12}$ respectively $10^{10}$ neurons. Hence one can ask the question if a learning rule based on a neuron memory can achieve comparable good results in learning as a learning rule based on synaptic counters \cite{cb1999,kbs2000,bc2001}. The insights one can obtain by answering this question are twofold. Firstly, by introduction of a learning rule based on neuron memory instead of synaptic memory one can show that the learning rules proposed by \cite{cb1999,kbs2000,bc2001} are not minimal in terms of economical use of resources. Secondly, a biological interpretation of the working mechanism of a learning rule with neuron memory could reveal novel insights of synaptic plasticity because the starting point was a more mathematical one. In the following we give a brief sketch of our way to a Hebb-like learning rule with neuron memory.

For the given topology of the neural network defined in section \ref{model_xor} as well as for any other network topology one can find a linear mapping $M$ with $c_n=Mc_s$. Here $c_s\in \mathds{N}^S$ and $c_n\in \mathds{N}^N$ are vectors which components are the synaptic and neuron counters. The components of the linear mapping $M$ are easily obtained by summing up the incoming synaptic counters of a neuron which has to be equal to the neuron counter of that neuron. The same holds for the sum of the synaptic counters which come out of the neuron \cite{vphd_2003}. 

The crucial point is that we do not use synaptic but neuron counters and hence we are interested in the inverse mapping. But for non trivial network topologies the number of synapses $S$ and neurons $N$ is different which gives a non quadratic matrix $M$ whose inverse is not defined by linear algebra. A way out of this is to use the Moore-Penrose pseudoinverse \cite{moore_1920,p1955} which is also defined for non quadratic matrices. Calculations for our three layer neural network reveal that in general the synaptic counters $c_{ij}$ are not simply the sum of the adjacent neuron counters $c_i$ and $c_j$ but also of far remote neuron counters \cite{vphd_2003}. This would result in a non local learning rule which violates the postulate of Hebb. To avoid this non-locality we introduce a stochastic instead of a deterministic approximation scheme which is described in the rest of this section.

Similar to \cite{cb1999,kbs2000,bc2001} again only active synapses $w_{ij}$ which were involved in the last signal processing step can be updated if $r=-1$ which correspond to a wrong network output. But now there is a stochastic condition
\begin{eqnarray}
p_\mathrm{coin}<p_\mathrm{\widetilde{c_{ij}}}^\mathrm{rank}
\end{eqnarray}
defined in \ref{p_rank} and \ref{p_coin} below which has to be fulfilled to update the synaptic weights by
\begin{eqnarray}
\label{synw1}
w_{ij}{\rightarrow}w'_{ij}=w_{ij}-{\delta},
\end{eqnarray}
with ${\delta}(t)={\delta}\in\mathds{R^+}$.

In addition to the network dynamics of the neurons there is a dynamics for the neuron counters $c_{i}$ which is defined by
\begin{eqnarray}
\label{synmem2}
c_{i}{\rightarrow}c`_{i}=\left\{ \begin{array}{r@{\quad}l}
{\Theta}, if &c_{i}-r>{\Theta}   \\c_{i}-r, if & {\Theta}\ge c_{i}-r\ge 0\\ 0, if & 0>c_{i}-r.
\end{array}\right.
\end{eqnarray}
Here ${\Theta}\in\mathds{N}$ is a threshold, $r=\pm1$ a reinforcement signal and $c_{i}$ a neuron counter. Equation \ref{synmem2} concerns only the neuron counters for the active neurons. The other neuron counters remain unchanged.  

To obtain the stochastic update condition $p_\mathrm{coin}<p_\mathrm{\widetilde{c_{ij}}}^\mathrm{rank}$ one has to follow through the following procedure:
\begin{enumerate}
\item Calculate the approximated synaptic counters $\widetilde{c_{ij}}$ of the active synapses by the neuron counters \ref{synmem2},
\begin{eqnarray}
\widetilde{c_{ij}}=c_i+c_j \label{approx_sync}
\end{eqnarray}
\item Because of $c_i\in\mathds{N}$ holds for all $i\in\{1,...,N\}$  $\Rightarrow\widetilde{c_{ij}}\in\mathds{N}$. Hence one can assign each approximated synaptic counter $\widetilde{c_{ij}}$ of each active synapse a probability $p_{\widetilde{c_{ij}}}^\mathrm{rank}$, which is given by the rank ordering distribution
\begin{eqnarray}
&P_{k}^\mathrm{rank} & \propto  k^{-\tau}  \label{p_rank}\\
&k  \in &\!\!\!\!\! \{1,\dots  ,2\Theta+3\} \label{p_rank_k}     \\   
&\tau  \in &\!\!\!\!\! \mathds{R}^+
\end{eqnarray}
with the mapping $k=2\Theta+3-\widetilde{c_{ij}}$ which is motivated by \cite{bp2001}. 
\item For the distribution $P(x)_\mathrm{coin}$ we also choose a power law 
\begin{eqnarray}
&P(x)_\mathrm{coin} &  \!\!\!\  \propto  x^{-\alpha}   \label{p_coin}\\
&x\in[0,1]  \\
&\alpha\in \mathds{R}^+
\end{eqnarray}
from which a probability $p_\mathrm{coin}$ is drawn for each active synapse.
\end{enumerate}

Let us compare and clarify the differences between the working mechanism of the learning rules proposed by \cite{cb1999,kbs2000} and ours. The learning rule by Bak and Chialvo updates all active synapses always if the reinforcement signal $r=-1$. The learning rule by Klemm, Bornholdt and Schuster updates an active synapse only if $r=-1$ and the synaptic counter exceeds a threshold. Our stochastic learning rule updates the active synapse only if $r=-1$ and the condition $p_\mathrm{coin}<p_\mathrm{\widetilde{c_{ij}}}^\mathrm{rank}$ is fulfilled that means with a certain probability which depends on the value of the approximated synaptic counter $\widetilde{c_{ij}}$.


\section{Results}\label{results}

We investigate the working mechanism of our novel stochastic Hebb-like learning rule with respect to the learning behavior of a three-layer feedforward network for the exclusive-or (XOR) problem and the influence of several different parameters, which constitute the entire system in table \ref{gen_sys}. We consider the influences of different network dynamics, the temperature-like parameter $\beta$, the noise $\eta$ and the three parameters $\Theta$, $\alpha$ and $\tau$ which determine our stochastic Hebb-like learning rule. The following subsections are subdivided according to the network dynamics. For all simulations we calculated the mean ensemble error $E(t)\in [0,1]$:
\begin{eqnarray}
E(t) & = & \frac{1}{N}\sum_{i=1}^Ne_i(t) \label{merror_gen}\\
e_i(t) & \in &  \{0,1\} \label{indiv_error}
\end{eqnarray}
to evaluate the network performance and hence its learning behavior. Here $e_i(t)\in \{0,1\}$ indicates if the output of network $i$ at time step $t$ was right $e_i(t)=0$ or wrong $e_i(t)=1$. The ensemble size is given in the respective subsections. In all simulations the synaptic weights $w_{ij}$ are i.i.d. initialized from the interval $[0,1]$ and the neuron counters $c_i$ are set to $0$.



\subsection{winner-take-all}\label{wta}

\begin{figure}[h!]
\centering
\epsfig{file=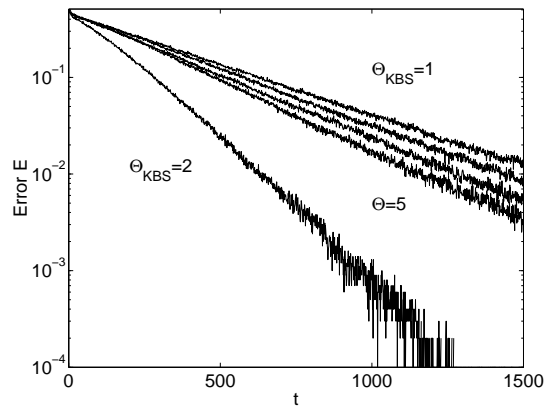, width=0.4\textwidth}
\caption{\label{v_wta_f1}Comparison of the mean error $E(t)$ for different values of $\Theta$. The notation ``KBS'' indicates the learning rule with a synaptic counter \cite{kbs2000}. For our learning rule with neuron counter $E$ is shown in dependence of $\Theta=5$, $\Theta=1$ and $\Theta=2$ (from bottom to top). The exponents $\tau$ and $\alpha$ are taken from table \ref{Emin_wta} at $t=1500$. The size of the ensemble was $10000$.}
\end{figure}

The convergence behavior of the winner-take-all mechanism as network dynamics is shown in figure \ref{v_wta_f1}. The mean ensemble error $E(t)$ is plotted for various $\Theta$ in a semi logarithmic plot. $E(t)$ converges rapidly for all $\Theta\in\{0,\ldots,5\}$ (only $\Theta\in \{5,1,2\}$ (from bottom to top are shown) within $1500$ time steps to an error below $10^{-2}$. 
\begin{figure*}[t!]
\begin{minipage}[c]{0.28\textwidth}
\epsfig{file=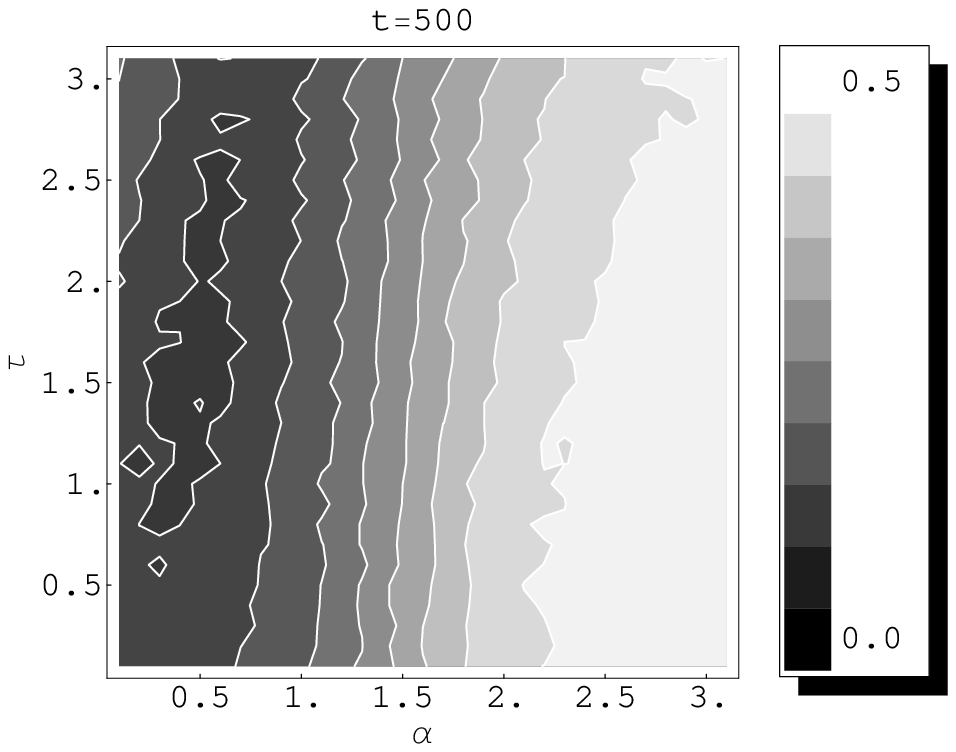, width=53mm}
\end{minipage}
\begin{minipage}[c]{0.28\textwidth}
\epsfig{file=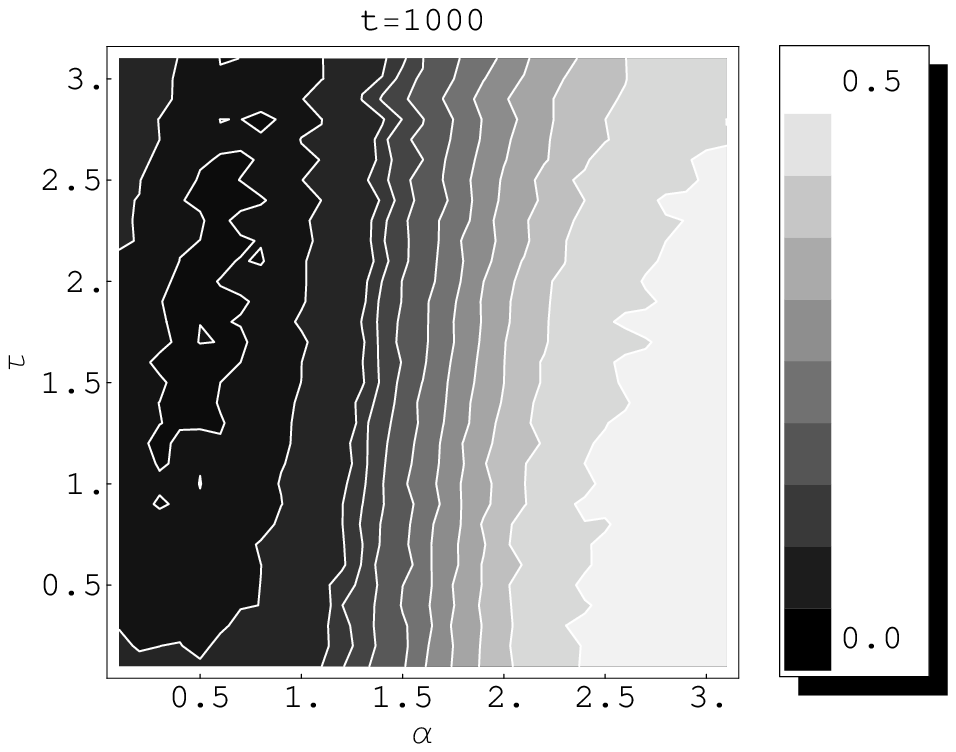, width=53mm}
\end{minipage}
\begin{minipage}[c]{0.28\textwidth}
\epsfig{file=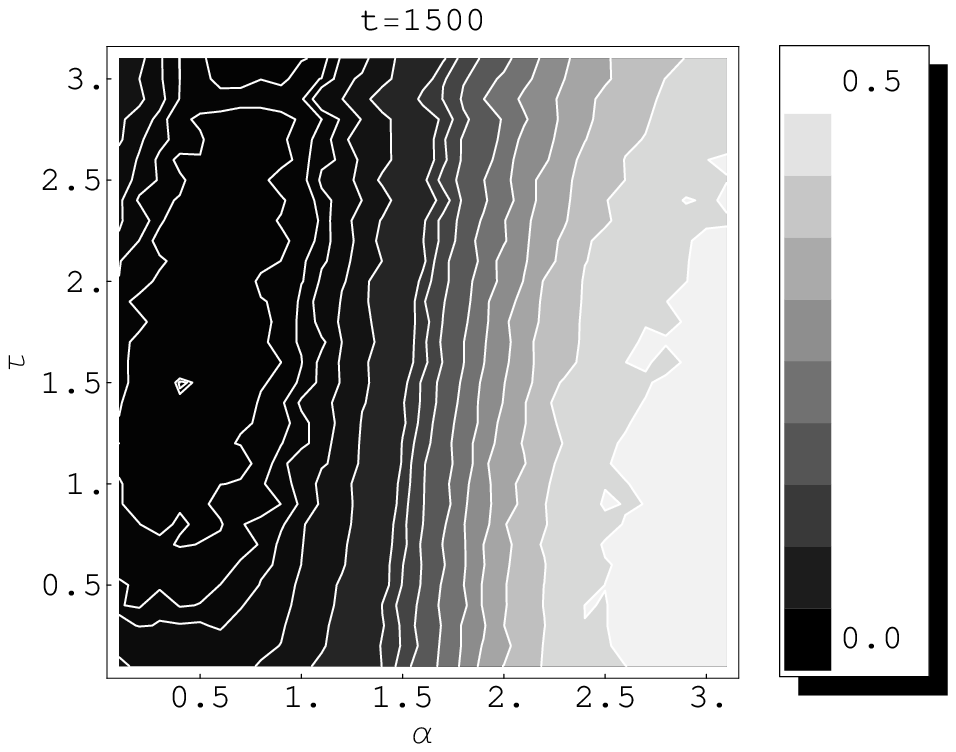, width=53mm}
\end{minipage}
\begin{minipage}[c]{0.28\textwidth}
\epsfig{file=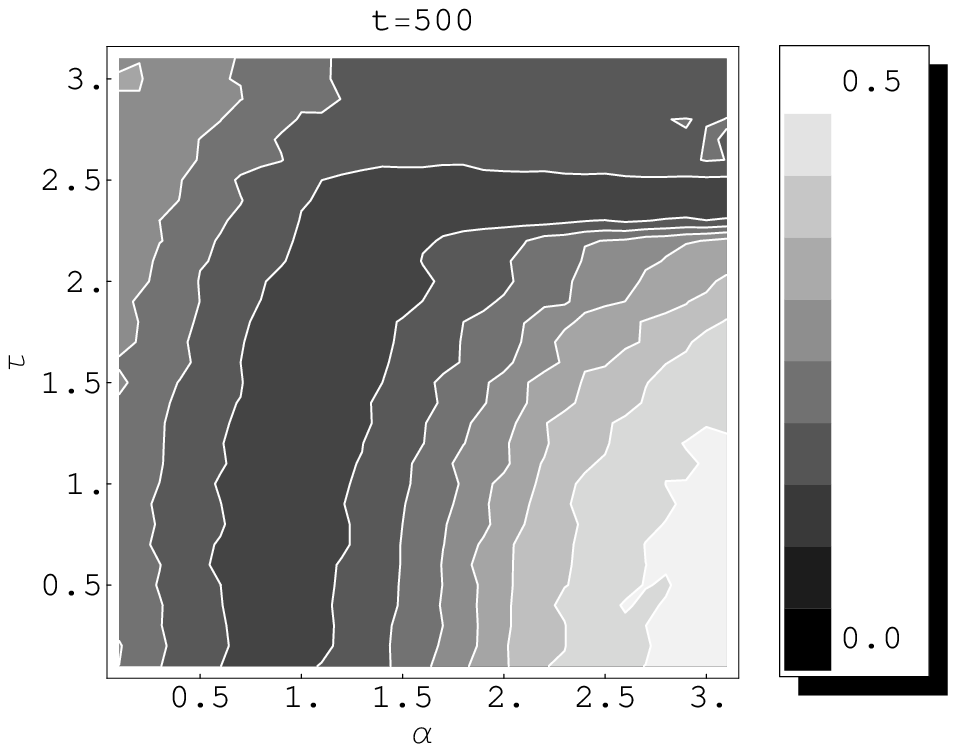, width=53mm}
\end{minipage}
\begin{minipage}[c]{0.28\textwidth}
\epsfig{file=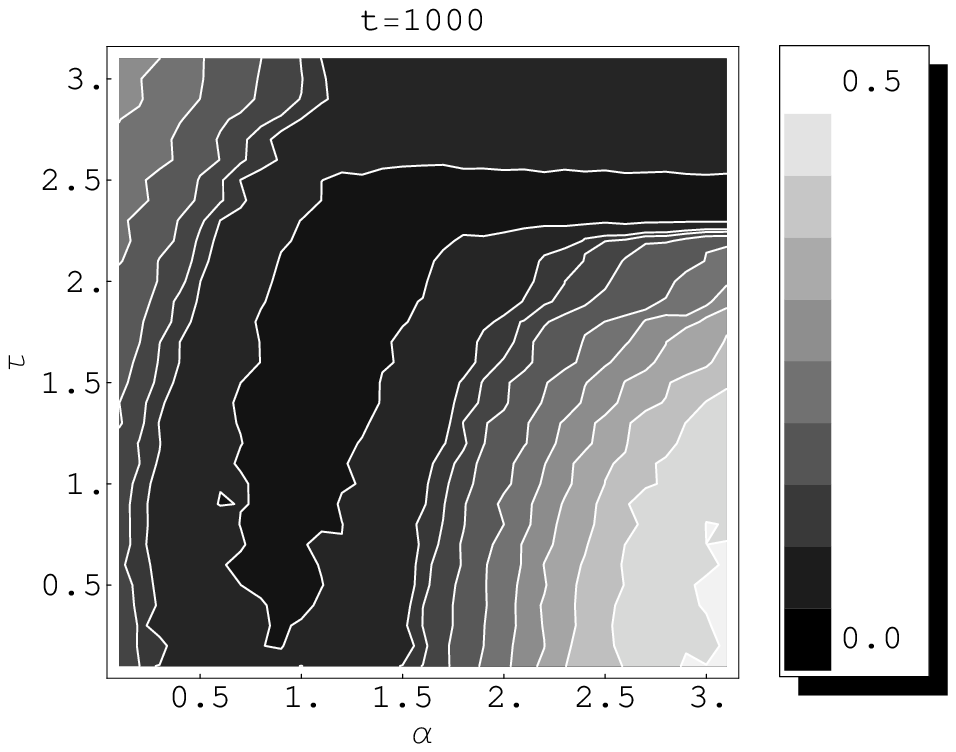, width=53mm}
\end{minipage}
\begin{minipage}[c]{0.28\textwidth}
\epsfig{file=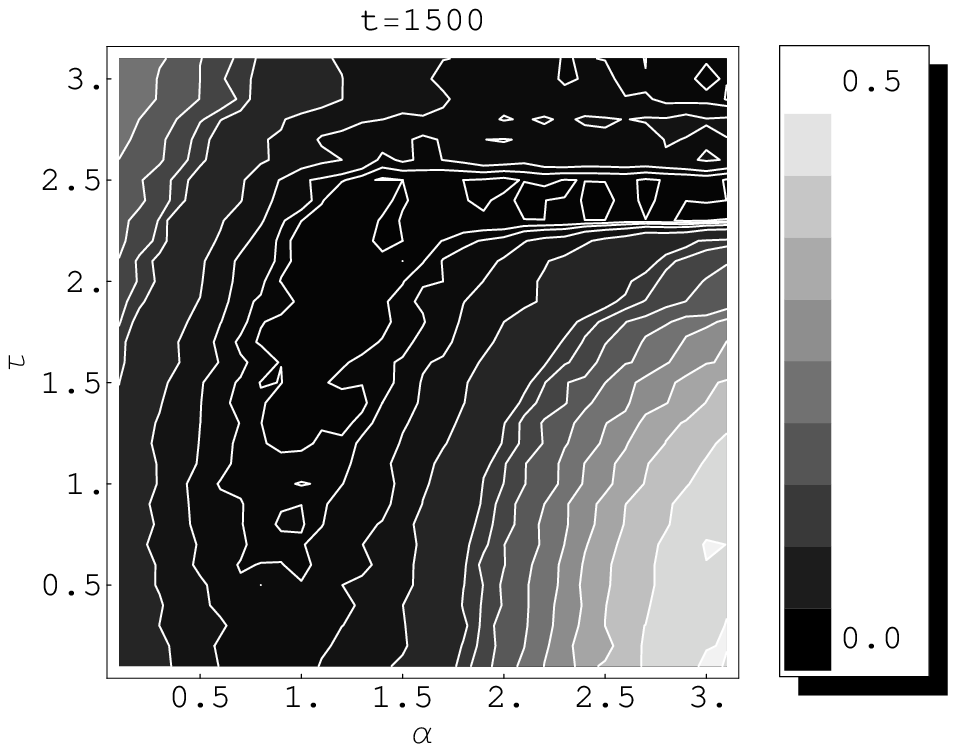, width=53mm}
\end{minipage}
\caption{\label{pw1}Mean error $E$ in dependence of $\tau$ and $\alpha$ at the time steps $t=500,1000$ and  $t=1500$ for $\Theta=0$ (top) and $\Theta=3$ (bottom). The network dynamics was a winner-take-all mechanism and the ensemble size $10000$.}
\end{figure*}

The best results are obtained for $\Theta=5$. A further increase of $\Theta$ does not improve the convergence behavior (not shown).
A comparison with the proposed learning rule by \cite{kbs2000} indicated by $\Theta_{\mathrm{KBS}}$ in figure \ref{v_wta_f1} reveals a stronger dependence of the synaptic memory with respect to the convergence behavior. This is a hint that our learning rule is due to its stochastic character regulated by the exponents $\alpha$ and $\tau$ more flexible with respect to different neuron memories which correspond to an evaluation of an individual failure rate as explained in \cite{kbs2000}.

\begin{table}[b!]
\caption{\label{Emin_wta}Minimal mean ensemble error $E_\mathrm{min}$ in dependence of the exponents $\alpha$ and $\tau$ and of the neuron counter $\Theta$, for the time steps $t=500,1000$ and $1500$ (left, middle and right column). (*) means that in this case there are three pairs of exponents for which the mean error is minimal. The other two pairs are: $\tau=2.7/3.0$, $\alpha=0.9/1.0$. The network dynamics was a winner-take-all mechanism and the ensemble size $10000$.}
\begin{ruledtabular}
\begin{tabular}{c|ccc|ccc|ccc} \hline
$\Theta$&&$\tau$&&&$\alpha$&&&$E_\mathrm{min}$&\\ \hline
0 &2.0&1.7&1.4&0.4&0.3&0.3&0.110&0.024&0.005 \\
1 &2.1&2.0&2.2&0.6&0.6&0.8&0.099&0.021&0.004  \\
2 &1.8&2.1&2.6&0.9&1.1&1.2&0.113&0.122&0.007  \\
3 &2.4&2.3&2.3&1.7&1.7&1.7&0.122&0.032&0.009 \\
4 &2.2&2.2&2.2&2.8&3.0&3.0&0.112&0.022&0.004  \\
5 &1.9&1.9&1.9&2.4&3.0&3.0&0.087&0.018&0.003 \\\hline
\end{tabular}
\end{ruledtabular}
\end{table}
The best exponents for the neuron memory of our stochastic learning rule were obtained by simulations in which we investigated systematically the dependence of the mean ensemble error $E$ of $\alpha$ and $\tau$. We chose the exponents from $0\le \alpha,\tau \le 3$ in discrete steps of $10^{-1}$ and let them fixed for the ensemble simulation. The corresponding results for $\Theta=0$ and $\Theta=3$ are shown in figure \ref{pw1}. In these gray scale contour plots the mean ensemble error $E$ is shown at three different time steps $t\in\{500,1000,1500\}$ which are chosen according to figure \ref{v_wta_f1}. Black corresponds to $E=0.0$ and white to $E=0.5$. One can see that there are regions in every plot which differ greatly in the respective mean ensemble error. Further more, the structure of the plots is for different $\Theta$ recognizable deformed. This can be qualitatively understood if one starts from $\Theta=0$ and $\alpha$ and $\tau$ values which correspond to $E_{\mathrm{min}}$. It follows from these assumptions that the update probability for fulfilling the condition $p_\mathrm{coin}<p_\mathrm{\widetilde{c_{ij}}}^\mathrm{rank}$ is optimal for these parameters. An increase in $\Theta$ leads to a decrease in $p_\mathrm{\widetilde{c_{ij}}}^\mathrm{rank}$ for each $\widetilde{c_{ij}}$ hence the update condition $p_\mathrm{coin}<p_\mathrm{\widetilde{c_{ij}}}^\mathrm{rank}$ is less often fulfilled. This can be compensated by an increase of $\alpha$ which reduces the average value of $p_\mathrm{coin}$ and so increases the probability that the update condition is satisfied. The interplay between $\Theta$, $\alpha$ and $\tau$ which constitute our stochastic learning rule regulates the synaptic update probability and hence results in the deformation observed in figure \ref{pw1}.

Table \ref{Emin_wta} gives a summary of the simulation results for all $\Theta\in\{0,\dots,5\}$ and shows the best exponents for which the mean ensemble error is minimal to the corresponding time steps $t\in\{500,1000,1500\}$. The most interesting result in table \ref{Emin_wta} is that the best $\alpha$ values are all greater than zero. This eliminates a equal distribution for $P(x)_\mathrm{coin}$ and provides additional justification for the power law ansatz in \ref{p_coin}.



\subsection{softmax}\label{softmax}

\begin{figure}[h!]
\epsfig{file=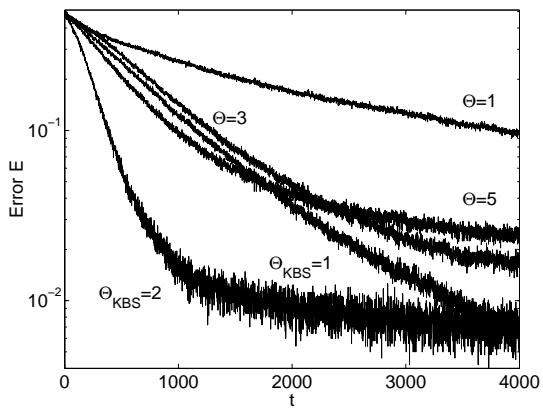, width=0.4\textwidth}
\caption{\label{v_b10_f1cp}Mean ensemble error $E$ versus time $t$ for several values of $\Theta$. The notation ``KBS'' again indicates that the learning rule with synaptic counters was used. The exponents $\tau$ and $\alpha$ are for the respective values of $\Theta$ taken from table \ref{Emin_b} for $t=1500$. All simulations use the softmax mechanism with $\beta=10$ as network dynamics. The size of the ensemble was $10000$.}
\end{figure}

We present now the results for a softmax mechanism as network dynamics for which the temperature-like parameter was chosen to $\beta=10$ (c.f. \cite{kbs2000}). The convergence behavior of the mean ensemble error $E$ is depicted in figure \ref{v_b10_f1cp}. Again learning takes place for all $\Theta$. But now one can see the formation of two different parameter groups. The first group with $\Theta=\{0,1,2\}$ shows a slower convergence than the second one and the curves for the mean error $E(t)$ are almost identical for the three $\Theta$ values. For this only $\Theta=1$ is shown. 
\begin{figure*}[t!]
\begin{minipage}[c]{0.28\textwidth}
\epsfig{file=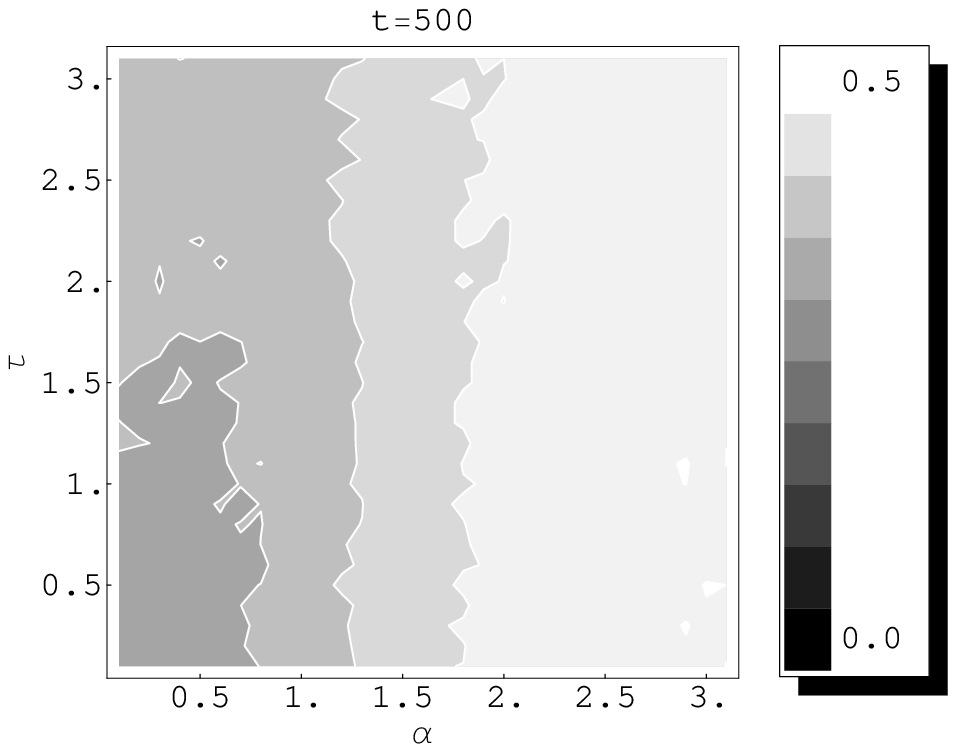, width=53mm}
\end{minipage}
\begin{minipage}[c]{0.28\textwidth}
\epsfig{file=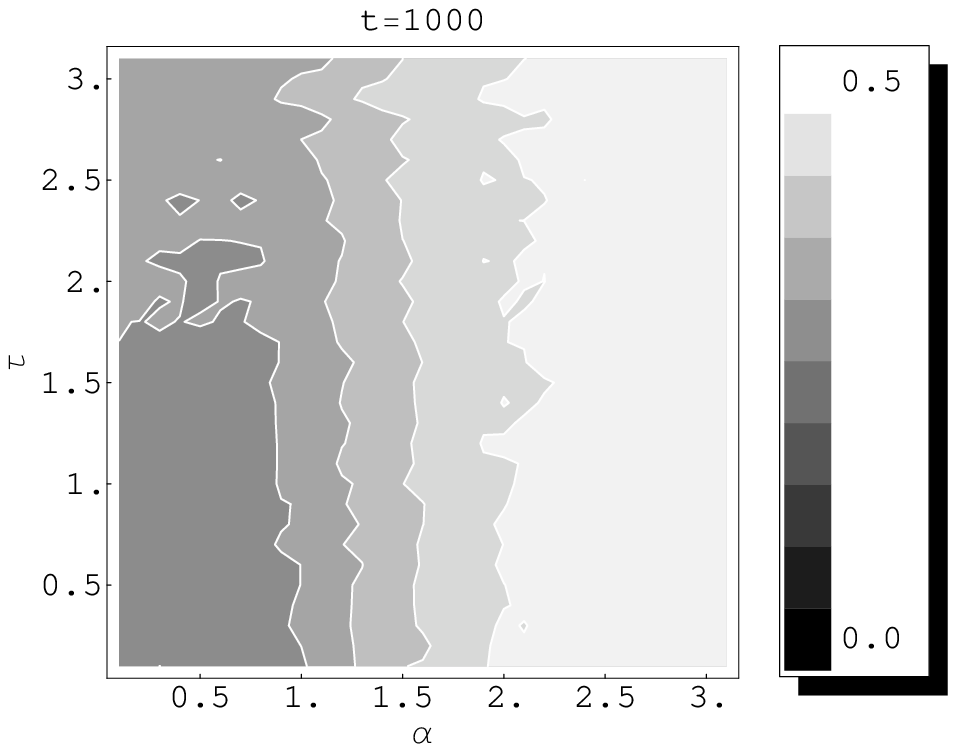, width=53mm}
\end{minipage}
\begin{minipage}[c]{0.28\textwidth}
\epsfig{file=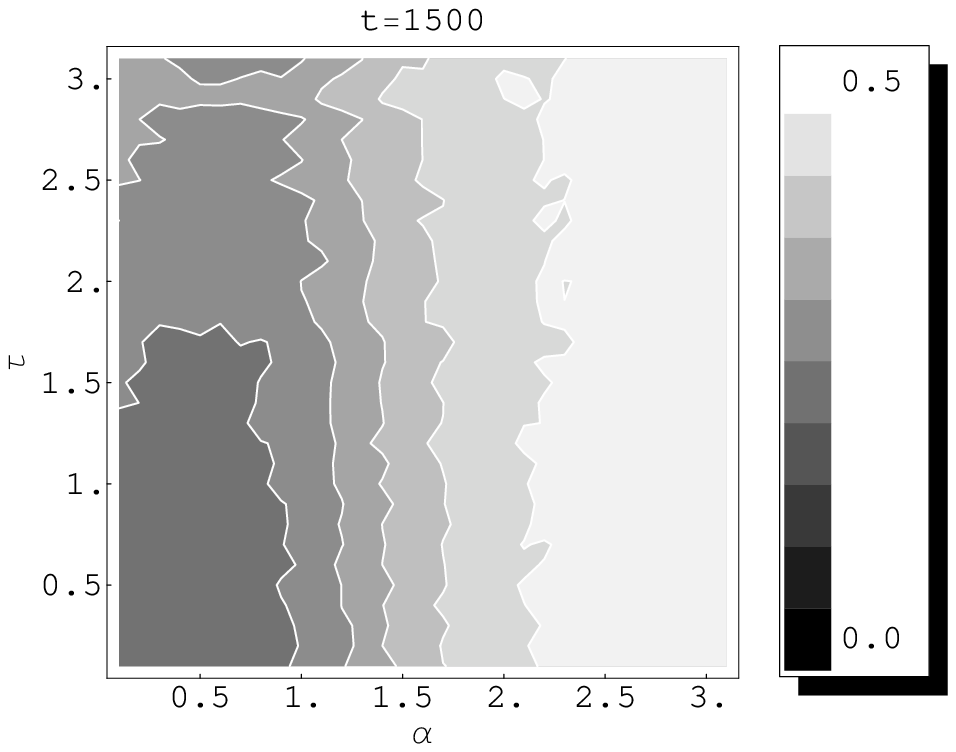, width=53mm}
\end{minipage}
\begin{minipage}[c]{0.28\textwidth}
\epsfig{file=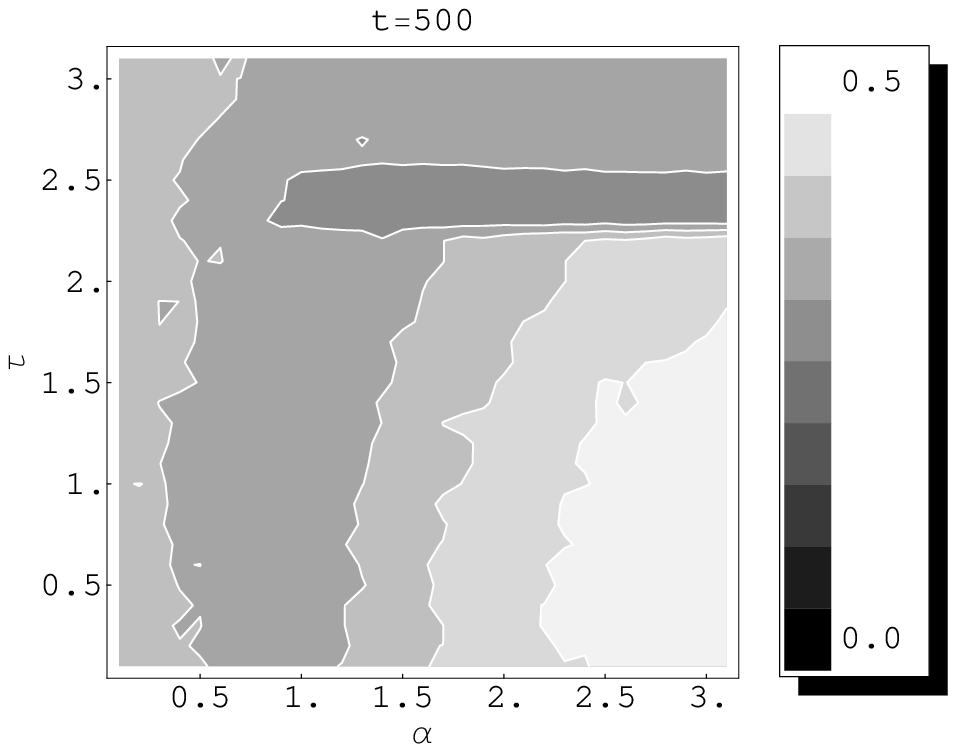, width=53mm}
\end{minipage}
\begin{minipage}[c]{0.28\textwidth}
\epsfig{file=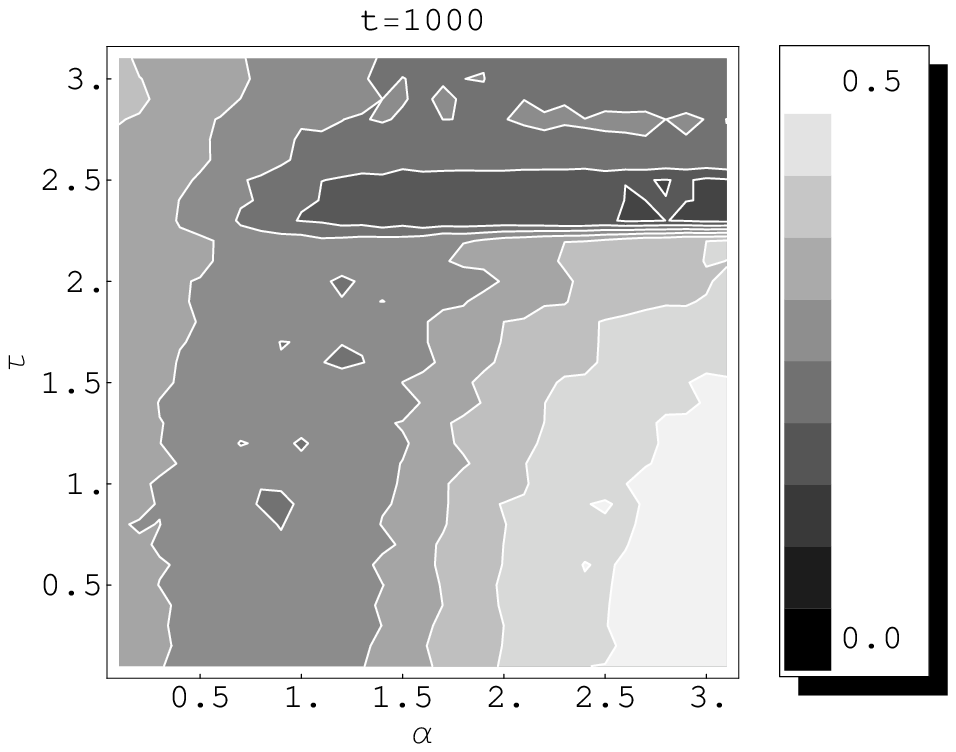, width=53mm}
\end{minipage}
\begin{minipage}[c]{0.28\textwidth}
\epsfig{file=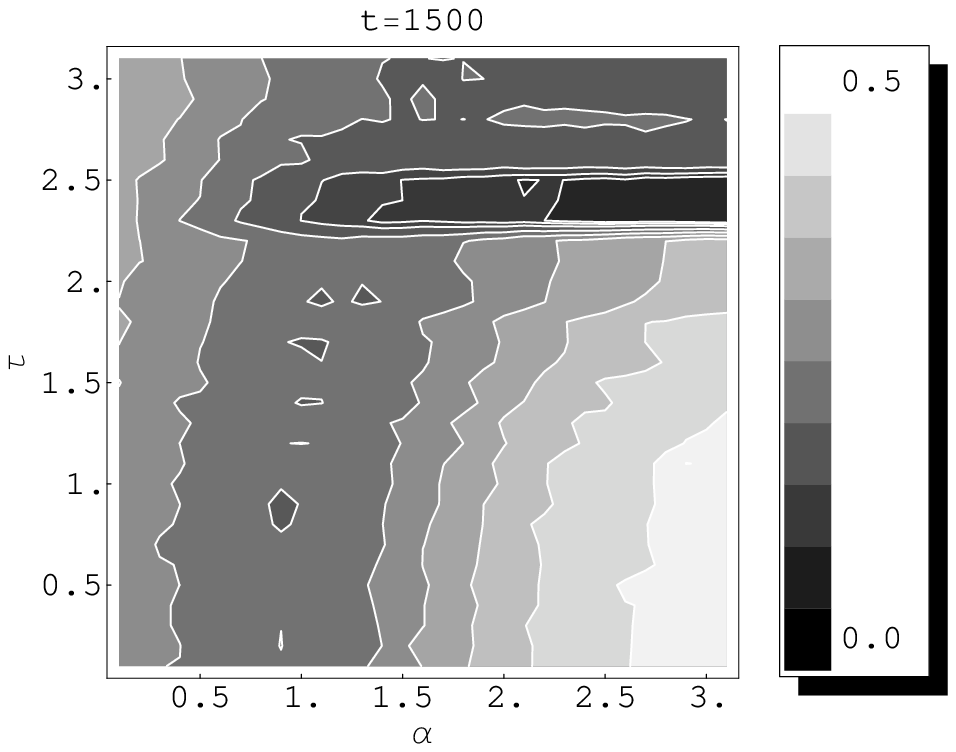, width=53mm}
\end{minipage}
\caption{\label{pw2}Mean ensemble error $E$ in dependence of $\tau$ and $\alpha$ at t=500, 1000 and 1500 for $\Theta=0$ (upper) and $\Theta=3$ (lower). The size of the ensemble was $10000$.}
\end{figure*}
The second group consisting of $\Theta=\{3,4,5\}$ is faster and can be clearly distinguished from the first group ($\Theta=4$ not shown is between $\Theta=3$ and $\Theta=5$). This is an effect of $\beta=10$ which introduces some kind of disorder in the system by the stochastic selection of an active neuron given by \ref{swta} and \ref{swtan}. In the case of a winner-take-all dynamics the selection of an active neuron is deterministic and not perturbed by the presence of a finite temperature or noise. Hence the average time to learn the XOR mapping is lower as in the case with finite temperature $\beta^{-1}$. This of course can be read from the values of the mean ensemble error $E(t)$ for corresponding $\Theta$ values by comparing the results in figure \ref{v_wta_f1} and \ref{v_b10_f1cp}. Thus there was no significant difference of the convergence time of different $\Theta$ values due to the flexibility of the learning rule regulated by the exponents $\alpha$ and $\tau$. For $\beta=10$ the limit of the flexibility of our stochastic learning rule is reached. Now there is a $\Theta$ below which the time used to average the individual failure rate of each synapse is to short for an effective learning process. Nevertheless even for $\Theta\le 2$ one can find adequate exponents $\alpha$ and $\tau$ for which learning takes place although significantly slower.

Comparing this with the learning rule of Klemm, Bornholdt and Schuster shows again that the best value for the synaptic counter $\Theta_{\mathrm{KBS}}=2$ converges fastest and reaches $E=10^{-2}$ at about $t=2000$. Note that in contrast to the winner-take-all mechanism shown in figure \ref{v_wta_f1} now we can observe intersections between different convergence curves however to times when $E$ is already below $10^{-1}$. 
\begin{table}[b!]
\caption{\label{Emin_b}Minimal mean error $E_\mathrm{min}$ in dependence of the exponents $\alpha$ and $\tau$ and of the neuron counter $\Theta$, for the time steps $t=500,1000$ and $1500$. The network dynamic was governed by the softmax mechanism with $\beta=10$.}
\begin{ruledtabular}
\begin{tabular}{c|ccc|ccc|ccc} \hline
$\Theta$&&$\tau$&&&$\alpha$&&&$E_\mathrm{min}$&\\ \hline
0 &0.2&0.0&0.5&0.3&0.2&0.1&0.325&0.250&0.200 \\
1 &1.8&1.0&0.5&0.5&0.6&0.4&0.314&0.246&0.200 \\
2 &3.0&3.0&3.0&1.2&1.1&1.3&0.312&0.237&0.197  \\
3 &2.4&2.2&2.2&2.0&3.0&1.3&0.264&0.138&0.075 \\
4 &2.2&2.2&2.2&3.0&2.9&3.0&0.239&0.121&0.069  \\
5 &1.9&1.9&1.9&2.9&3.0&2.7&0.199&0.088&0.055 \\\hline
\end{tabular}
\end{ruledtabular}
\end{table}

The exponents used in figure \ref{v_b10_f1cp} are again obtained by simulations for all values out of the interval $0\le \alpha,\tau \le 3$ in discrete steps of $10^{-1}$. The results analog to figure \ref{pw1} are shown in figure \ref{pw2}. One can see that the situation is quite similar to figure \ref{pw1} whereas the structure due to the influence of $\beta=10$ is now more succinct. A little surprising is the fact that the overall structure remained almost unchanged. Hence the influence of $\beta$ seems to be almost linear at least up to $\beta=10$. Table \ref{Emin_b} gives a summary of all simulation results and shows the best exponents for which the mean ensemble error $E$ is minimal to the corresponding time steps $t\in\{500,1000,1500\}$. Again $\alpha$ is always greater than zero which excludes an equal distribution for $P_{\mathrm{coin}}$.



\subsection{noisy winner-take-all}\label{nwta}

In this subsection we use the noisy winner-take-all mechanism \ref{nwta_c4}, \ref{nwtas_c4} as network dynamics. In contrast to the preceding results we investigate now the convergence behavior over long time scales under the influence of additive noise $\eta$ over some orders of magnitude. Further more we compare the influence of the size of the synaptic change $\delta$. For this we use $\delta=1=\mathrm{const.}$ like in the preceding simulations and compare this with $\delta$ uniformly drawn out of $[0,20]$. Again we did the simulations for all values $\Theta\in \{0,\dots,5\}$ but show only the results for $\Theta=0$ and $\Theta=5$ which give the significant differences.

The evaluation of the mean ensemble error $E$ is here a little different. We simulate as long as it takes the network to reach a stationary fix point and then average over the next $T_m$ time steps. In addition we average over a (small) ensemble $N$.
\begin{eqnarray}
&E&=\frac{1}{N(T_m+1)}\sum^{N}_{i=1}\sum^{T}_{t=T-T_m}e_i(t) \label{merror_conv1_long}\\
&e_i(t)&\in \{0,1\} \label{merror_conv2_long}
\end{eqnarray}
Here $N=100$, $T=100000$ and $T_m=10000$.

\begin{figure}[h!]
\begin{minipage}[c]{0.23\textwidth}
\epsfig{file=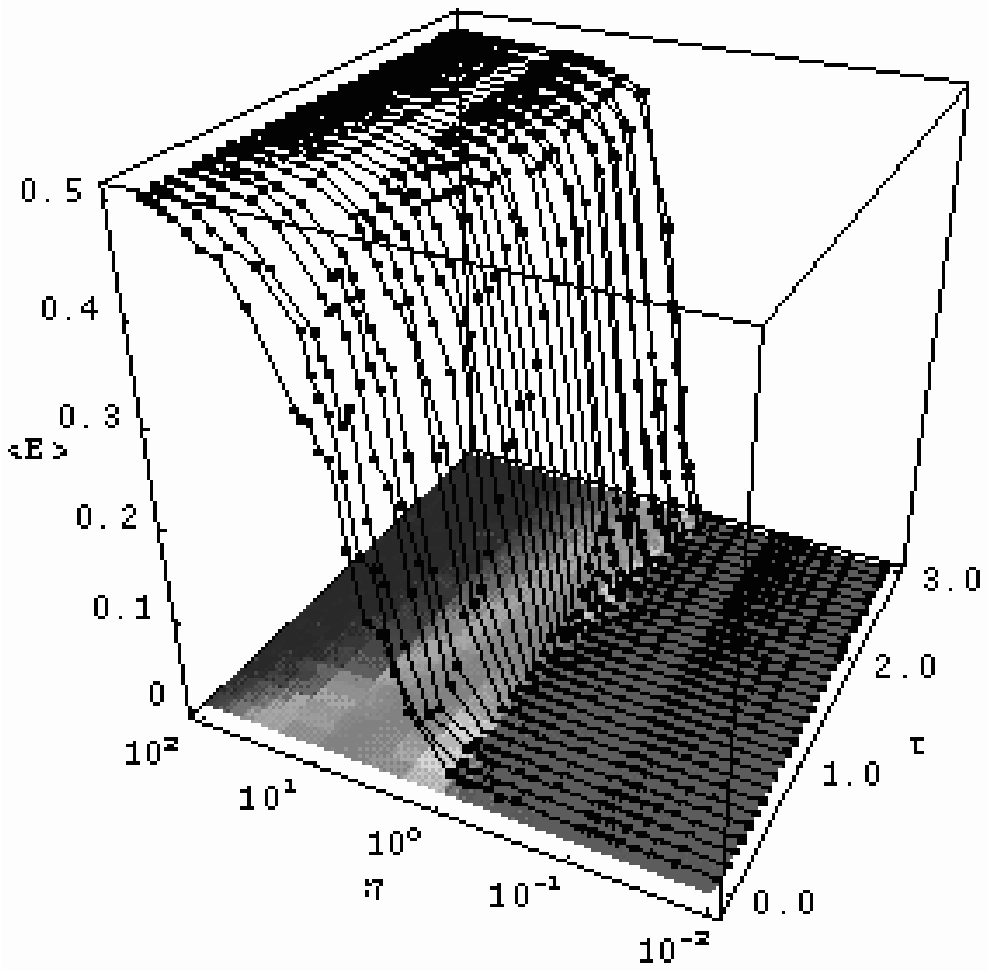, width=1.0\textwidth}
\end{minipage}
\begin{minipage}[c]{0.23\textwidth}
\epsfig{file=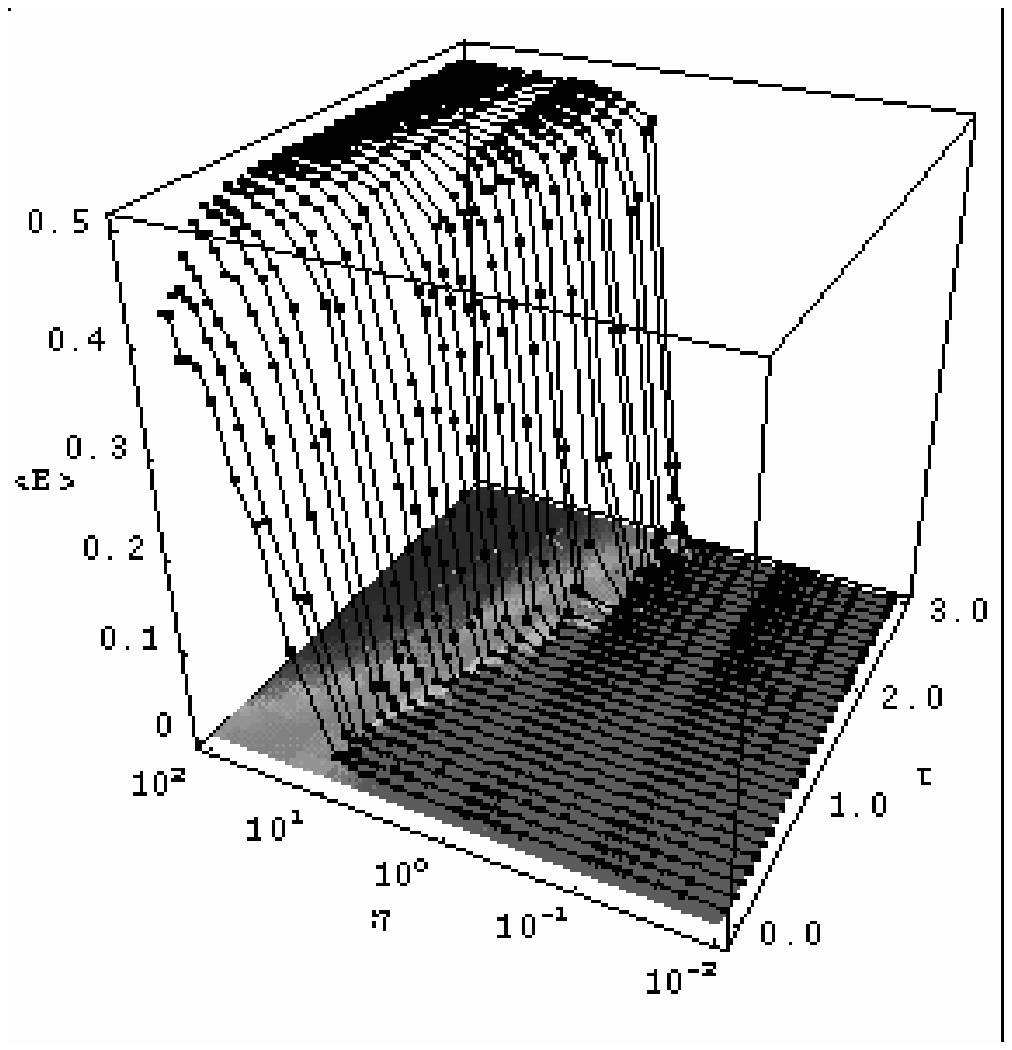, width=1.0\textwidth}
\end{minipage}
\begin{minipage}[c]{0.23\textwidth}
\epsfig{file=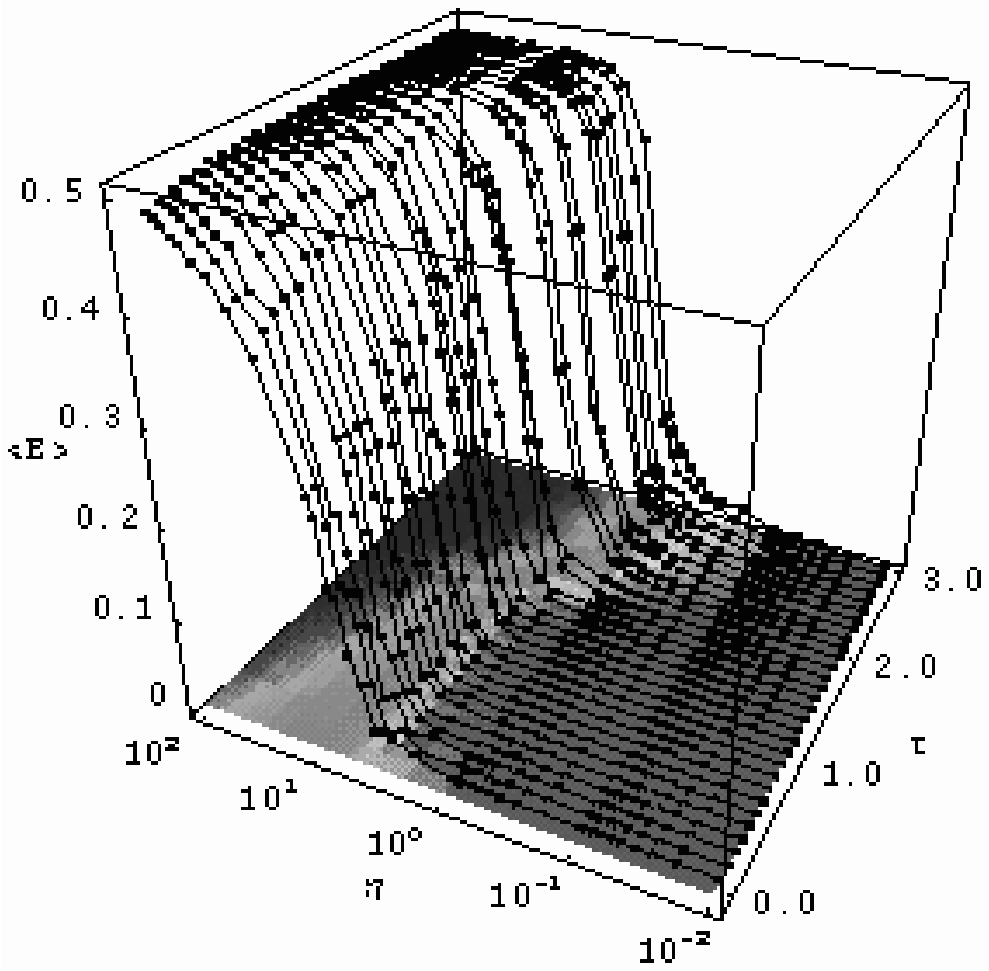, width=1.0\textwidth}
\end{minipage}
\begin{minipage}[c]{0.23\textwidth}
\epsfig{file=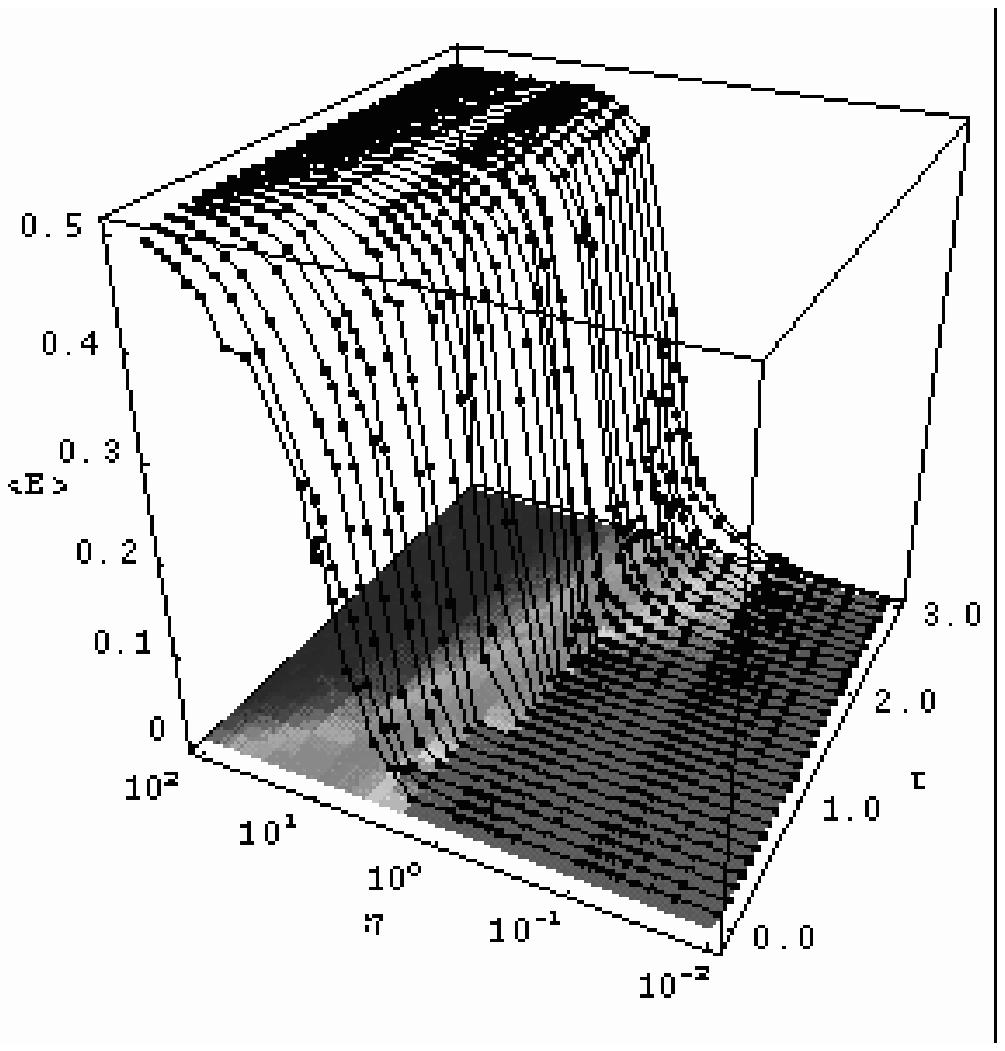, width=1.0\textwidth}
\end{minipage}
\caption{\label{pht1}Mean ensemble error $E$ in dependence of the noise $\eta$ and the exponent $\tau$. Here $\alpha=1.5$ was chosen fixed. Left: $\delta=1$. Right: $\delta\in[0,20]$. Top figure: $\Theta=0$. Bottom figure: $\Theta=5$.}
\end{figure}

\begin{figure}[h!]
\begin{minipage}[c]{0.23\textwidth}
\epsfig{file=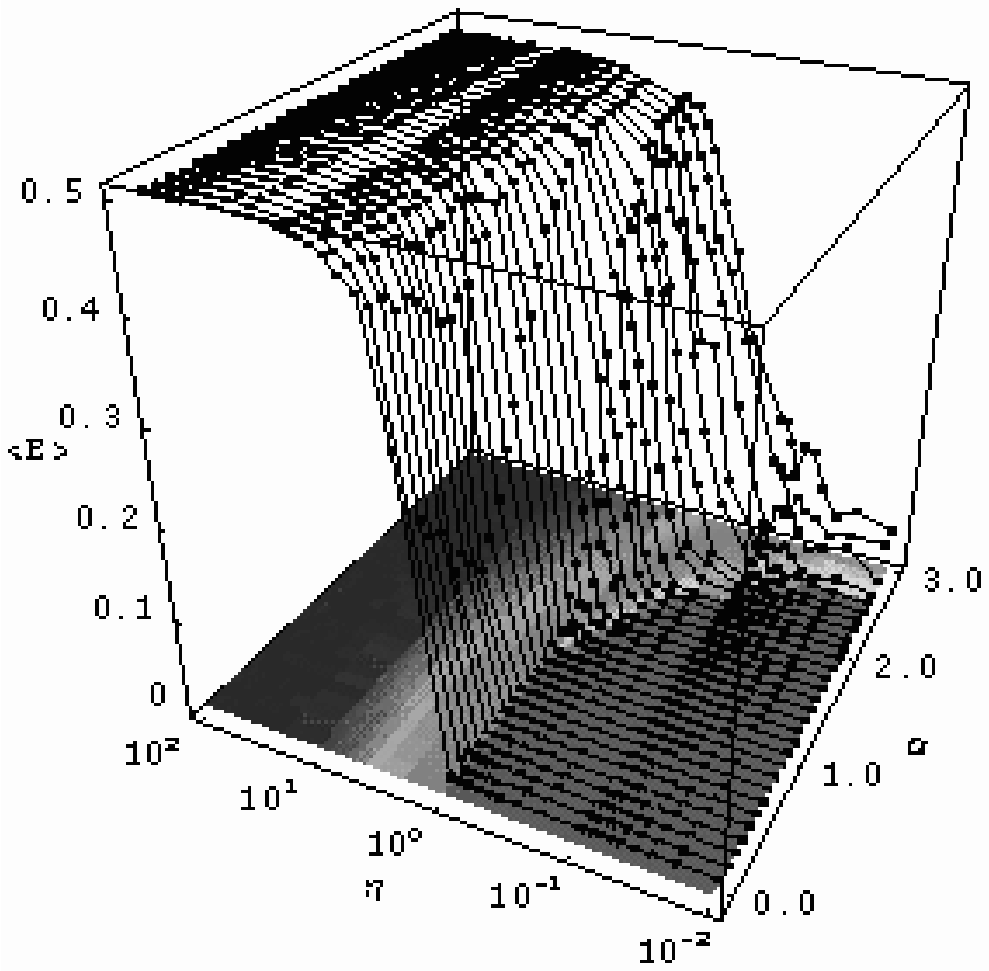, width=1.0\textwidth}
\end{minipage}
\begin{minipage}[c]{0.23\textwidth}
\epsfig{file=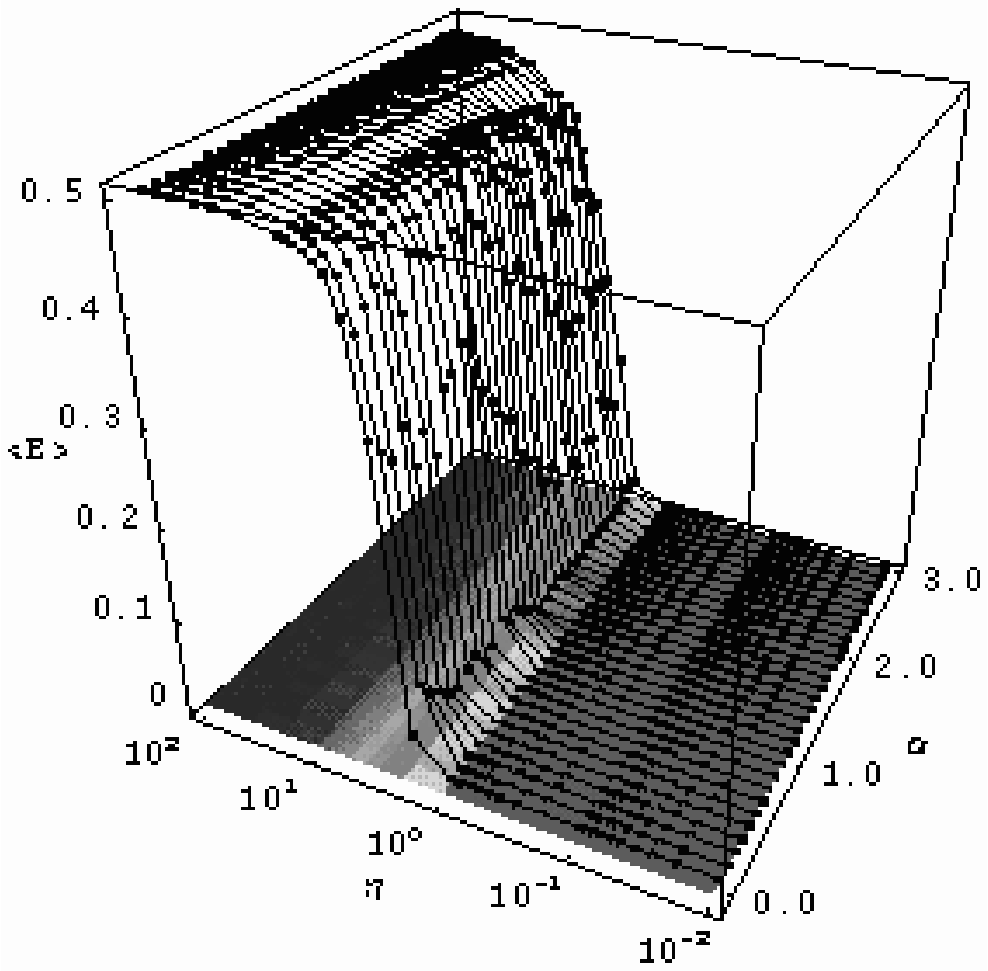, width=1.0\textwidth}
\end{minipage}
\begin{minipage}[c]{0.23\textwidth}
\epsfig{file=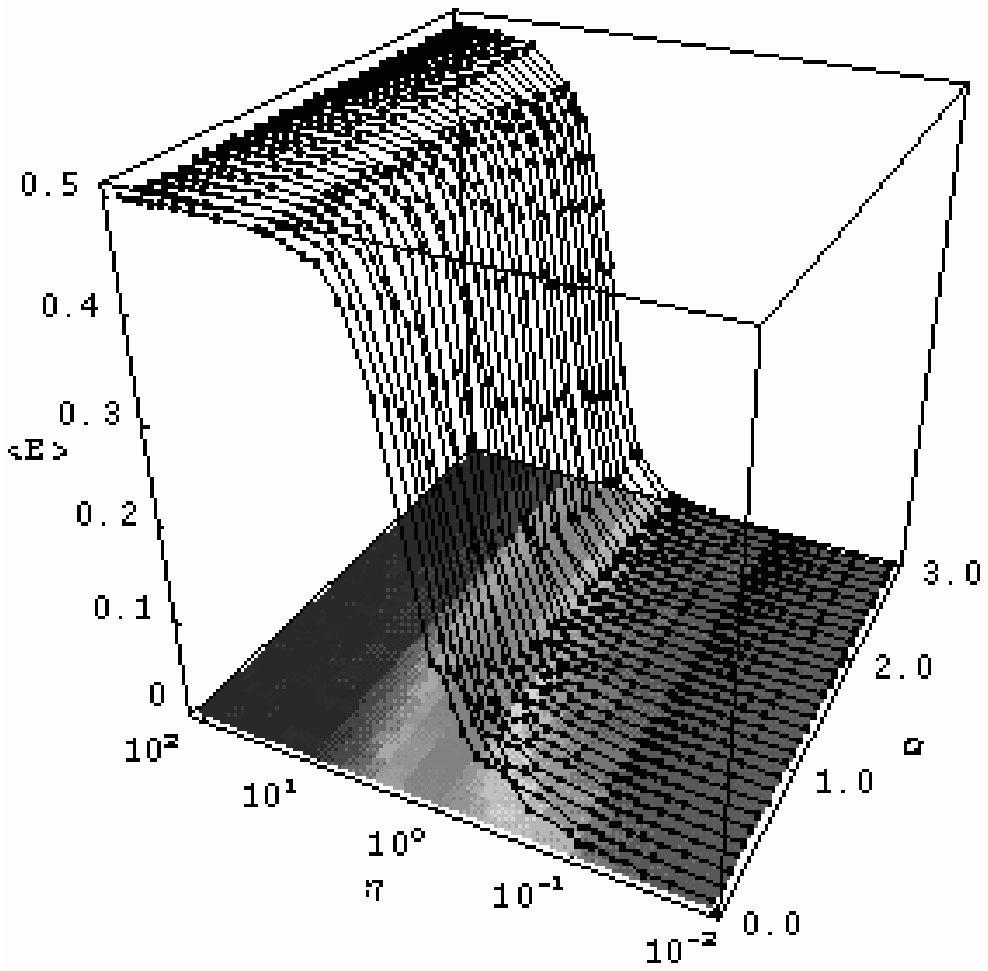, width=1.0\textwidth}
\end{minipage}
\begin{minipage}[c]{0.23\textwidth}
\epsfig{file=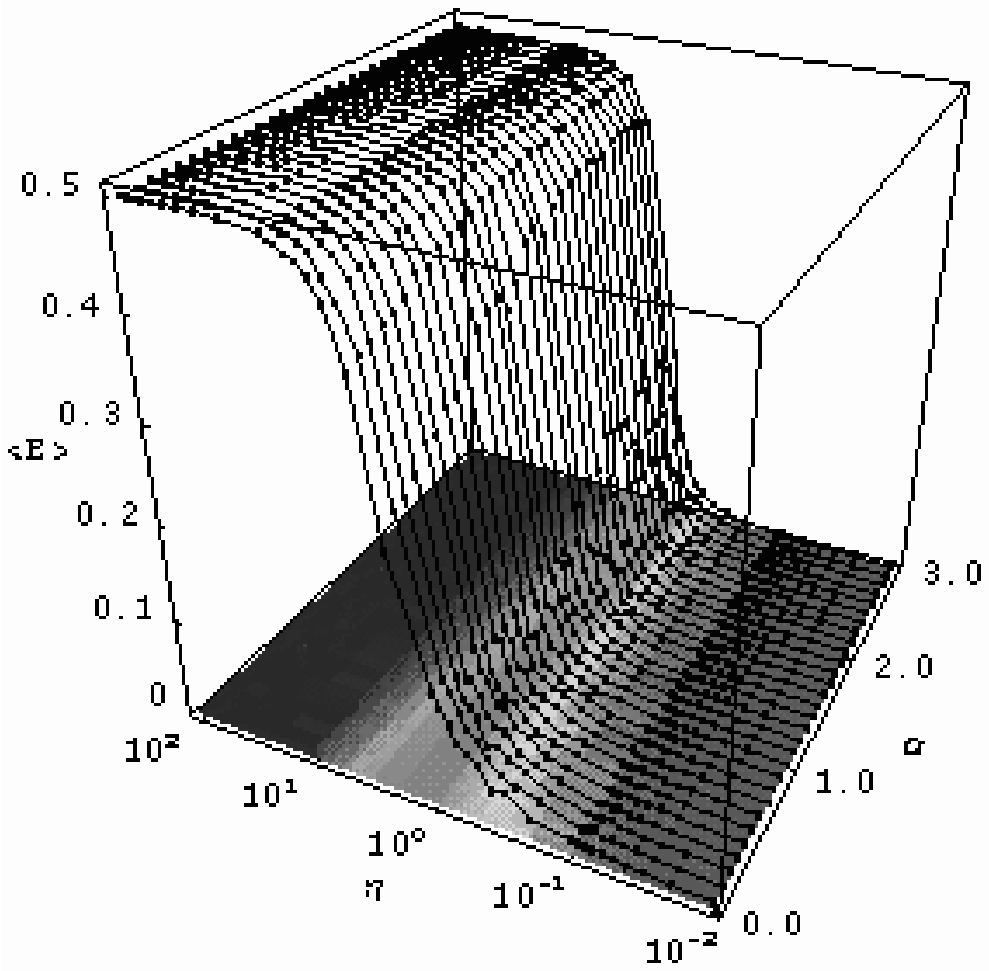, width=1.0\textwidth}
\end{minipage}
\caption{\label{pht2}Mean ensemble error $E$ in dependence of the noise $\eta$ and the exponent $\tau$. Here $\tau=2.0$ was chosen fixed. Left: $\delta=1$. Right: $\delta\in[0,20]$. Top figure: $\Theta=0$. Bottom figure: $\Theta=5$.}
\end{figure}

The first results are shown in figure \ref{pht1}. Here the mean ensemble error $E$ is investigated in dependence of the noise $\eta$ and the exponent $\tau$ of the rank ordering distribution. The exponent of the distribution $P(x)_\mathrm{coin}$ was fixed to $\alpha=1.5$ under regard of the preceding results. In figure \ref{pht1} as well as in figure \ref{pht2} comparable $\Theta$ values are arranged in rows with $\Theta=0$ in the top and $\Theta=5$ in the bottom row. The left columns give results for $\delta=1=\mathrm{const.}\sim O(1)$ and the right for $\delta\in [0,20]\sim O(10)$. From the scales of the  synaptic changes the steps occurring in figure \ref{pht1} and \ref{pht2} at noise values of order $\eta\sim 10^0$ respectively $\eta\sim 10^1$ are crudely explained.

In the first row of figure \ref{pht1} one can clearly see that the additional degree of freedom in form of a variable size of synaptic alterations $\delta$ leads to a significant improvement for $\tau<2.0$ of one scale of order in $\eta$. This effect can be gradually reduced by increasing $\Theta$ \cite{vphd_2003} from $0$ to $5$. The lower part of figure \ref{pht1} gives the final results for this procedure for $\Theta=5$. Here the influence of the additional degree of freedom is not only completely eliminated but one obtains for $\delta=1$ even better results in the range $1.0<\tau<2.0$.

This can be understood taking into consideration that learning is effective if the synaptic weights are changed as fast as possible and as often as necessary. On the first sting this looks like a contradiction but new paths in the network which connect input with output neurons correctly can be found only if synapses are changed. On the other hand by a synaptic change old paths in the neural network which are already learned correctly can be destroyed and hence unlearned. For this the three parameters constituting our learning rule $\Theta$, $\alpha$ and $\tau$ have to be chosen so that the probability to fulfill the update condition $p_\mathrm{coin}<p_\mathrm{\widetilde{c_{ij}}}^\mathrm{rank}$ is in accord with the motto given above.

Let us start at $\delta=1$ and $\Theta=0$ with a qualitative explanation. Here low $\tau$ values give either equal or slightly better results. This indicates that the probability to fulfill the update condition $p_\mathrm{coin}<p_\mathrm{\widetilde{c_{ij}}}^\mathrm{rank}$ is more adequate for low $\tau$ values than for high. For low $\tau$ values the update probability is less than for high $\tau$ values because of equation \ref{p_rank} under the natural assumption that the neural network did not learn the XOR problem up to a certain time step which implies high values of the approximated synaptic counters $\widetilde{c_{ij}}$. This holds for every $\Theta$ value \cite{vphd_2003} and hence also for $\Theta=5$ shown in the bottom left figure in \ref{pht1}. 

When $\Theta$ is gradually increased from $\Theta=0$ to $\Theta=5$ there is an additional effect on the performance. The update probability crosses the threshold from too high to too low values. This can be seen by increasing $\Theta$ for fixed $\tau$ values which causes a decrease of the update probability because of equation \ref{p_rank_k}. 
In the bottom left figure \ref{pht1} one can see that for $\Theta=5$ and $\tau<2.0$ this leads to an improvement of the performance of the network. This is in accordance with the explanation for the top left figure. For $\tau\ge2.0$ one would expect worse results than for $\tau<2.0$ which is true but better results for $\Theta=5$ than for $\Theta=0$. This however does not hold because for these parameters the update probability crossed the critical threshold and hence is too small which prevents an efficient learning.

If the update probability is greater or smaller than this critical threshold can also be seen from the steps occurring in figure \ref{pht1}. A step occurring at $\eta\sim O(1)$ is for $\delta=1\sim O(1)$ obviously. A shift to higher or lower noise values indicates a decreased (increased) update probability because averaging over the past network outcomes is prolonged (reduced). This follows from the fact that the average time over a time series which is contaminated by noise to detect a signal has to be longer the higher the influence of the noise is.

The influence of a variable synaptic change $\delta$ consists in an enhancement of the effects described above. An update probability which is too high is stronger punished due to higher mean synaptic changes and results in a higher probability to destroy already learned paths in the network. This can be seen in both of the right figures in  \ref{pht1}. 

Figure \ref{pht2} shows the same results for the mean ensemble error $E$ as figure \ref{pht1} but now $\alpha$ is variable and $\tau=2.0$ is constant. The occurring effects are again explained by the influence of the three parameters $\Theta$, $\alpha$ and $\tau$ on the update probability of the condition $p_\mathrm{coin}<p_\mathrm{\widetilde{c_{ij}}}^\mathrm{rank}$. The most interesting result here is the strong dependence of the mean ensemble error $E$ of $\Theta$ for $\delta=1$. For a value of $\Theta=0$ the exclusive-or (XOR) problem can not be learned completely for $\tau\sim 3.0$ even for very low noise values but only for low $\tau$ values. The situation is almost completely changed for $\Theta=5$. Now the performance for $\tau\sim 0.0$ is worse than for higher $\tau$ values. This reflects too high (low) update probabilities for $\Theta=0$ and $\tau\sim 3.0$ ($\Theta=5$ and $\tau\sim 0.0$).



\section{Biological interpretation}\label{biointerpretation}

\begin{figure}[h!]
\epsfig{file=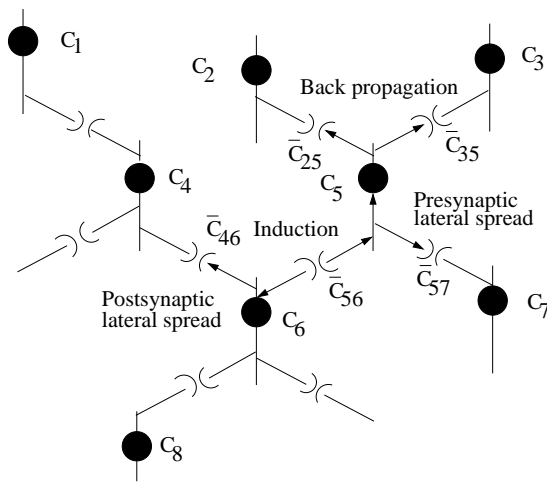, width=0.4\textwidth}
\caption{\label{network_heterosyn_ncounter}Schematic depiction of the interplay of the neuron counters and their influence on the approximated synaptic counters. $c_i$ are the neuron counters and $\tilde{c}_{ij}$ are the approximated synaptic counters.}
\end{figure}

In recent years there is an increasing number of experimental results which investigate heterosynaptic plasticity. In contrast to homosynaptic plasticity where only the synapse between active pre- and postsynaptic neuron is changed in form of either {\it{long-term depression}} (LTD) or {\it{long-term potentiation}} (LTP) heterosynaptic plasticity concerns also further remote synapses of the pre- and postsynaptic neuron. This scenario is schematically depicted in figure \ref{network_heterosyn_ncounter}. There we suppose neuron 5 and 6 were active and induced (homo-)synaptic plasticity on the synapse which is enclosed by these neurons. In addition to this form of plasticity Fitzsimonds et al. \cite{fitzsimonds_1997} found in cultured hippocampal neurons that the induction of LTD is also accompanied by back propagation of depression in the dendrite tree of the presynaptic neuron. Further more, depression also propagates laterally in the pre- and postsynaptic neuron. Similar results hold for the propagation of LTP, see \cite{bipoo_2001} for a review.

The correspondence to our learning rule follows immediately from the working principle of our neuron counters. In figure \ref{network_heterosyn_ncounter} the neuron counters are shown as $c_i$, $i\in \{1,\dots,8\}$ for each neuron in the schematic network. 
According to our learning rule there is a communication between the neuron counters of adjacent neurons. This communication leads to the formation of the approximated synaptic counters $\tilde{c}_{ij}$. From this, one can see that an alteration of the neuron counters $c_5$ and $c_6$ leads not only to an alteration of $\tilde{c}_{56}$ but also of all approximated synaptic counters $\tilde{c}_{k5}$, $\tilde{c}_{5l}$, $\tilde{c}_{m6}$ and $\tilde{c}_{6n}$ with $k\in\{2,3\}$, $l\in\{6,7\}$, $m\in\{4,5\}$ and $n\in\{8\}$. In biological terms $\tilde{c}_{k5}$  corresponds to backpropagation, $\tilde{c}_{5l}$ to presynaptic lateral and $\tilde{c}_{m6}$ to postsynaptic lateral spread of LTD. Interestingly the term $\tilde{c}_{6n}$ which would correspond to forward propagated postsynaptic LTD was not experimentally found up to now \cite{bipoo_2001}. 

A biological explanation for the cellular mechanisms of these findings are currently under investigation. Fitzsimonds et. al. \cite{fitzsimonds_1997} suggest the existence of retrograde signaling from the post- to the presynaptic neuron which could produce a secondary cytoplasmic factor for back-propagation and presynaptic lateral spread of LTD. On the postsynaptic side lateral spread of LTD could be explained similarly under the assumption that there is a blocking mechanism for the cytoplasmic factor which prevents forward propagated LTD. They are of the opinion that extracellular diffusible factors are of minor importance. In an abstract sense the approximated synaptic counters of our learning rule could be interpreted as an intracellular mechanism and not as an extracellular one. This would be consistent with the suggestions of \cite{fitzsimonds_1997}.

The future will show if further experiments confirm or reject the non-existence of forward propagated LTD. From a theoretical point of view and based on the assumptions made in this paper such a symmetry breaking mechanism occurring during the propagation of heterosynaptic LTD would be more elaborated than our stochastic Hebb-like learning rule.


\section{Conclusions}\label{conclusions}

In this article we presented a novel stochastic Hebb-like learning rule for neural networks and demonstrated its working mechanism exemplary in learning the exclusive-or (XOR) problem in a three-layer network. We investigated the convergence behavior by extensive numerical simulations in dependence of three different network dynamics which correspond all to biological forms of lateral inhibition. We found in all cases, parameter configurations for $\Theta$, the length of the neuron memory, $\alpha$, the exponent of the coin distribution and $\tau$, the exponent of the rank ordering distribution, which constitute the Hebb-like learning rule, to obtain not only a solution to the exclusive-or (XOR) problem but comparably well results to the learning rule recently proposed by Klemm, Bornholdt and Schuster \cite{kbs2000}. Comparably well means that for the exclusive-or (XOR) problem $\Theta_{\mathrm{KBS}}=2$ was always better than any parameter configuration $\{\Theta,\alpha,\tau\}$ for our learning rule, but for $\Theta_{\mathrm{KBS}}\not=2$ there are a lot of parameter configurations $\{\Theta,\alpha,\tau\}$ which result in a faster convergence in dependence of the time scale. In this point we agree with \cite{simon_1996,bc2001} where they take the opinion that natural systems try to solve problems satisficing and not optimally in a mathematical sense because of the lack of information biological systems are faced due to their inherent open character. In this respect our model consists of a large variety of parameters which work similar well without the need to find the very best parameter configuration. This parameter configuration can of course be found as shown in section \ref{wta} and \ref{softmax}. But that does not mean that other parameter configurations does not work at all. Our aim was to establish a Hebb-like learning rule which is very flexible with respect to special choices of the three parameters $\{\Theta,\alpha,\tau\}$.

Moreover our learning rule works comparably well to \cite{kbs2000} if one keeps in mind that our learning rule uses much less parameters. Because the number of neurons is always (much) less then the number of synapses the same holds for the respective numbers of synaptic and neuron counters which were used in the learning rules.

An interesting implication of our learning rule and its inherent stochastic character is that it offers a very simple qualitatively explanation of heterosynaptic plasticity which is observed experimentally. In addition to the experimentally observed back-propagation, pre- and postsynaptic lateral spread of {\em{long-term depression}} (LTD) our learning rule predicts forward propagated postsynaptic LTD for reasons of a symmetric communication between adjacent neurons. As far as we know there is no theoretical explanation of that phenomenon so far.

In further investigations we will demonstrate that our learning rule is not restricted to a multilayer network topology but works also in a class of recurrent networks constructed by an algorithm of Watts and Strogatz \cite{WattsStro_1998} when learning the problem of timing \cite{vhebb3_2003}. Moreover, it would be of interest to enlighten the power law ansatz for the rank ordering \ref{p_rank} and coin distribution \ref{p_coin} which was motivated by \cite{bp2001} in a more general context of stochastic optimization methods for rule-based systems.


\begin{acknowledgements}
We would like to thank Rolf D. Henkel, Klaus Pawelzik and Roland Rothenstein for fruitful discussions and Tom Bielefeld for carefully reading the manuscript.
\end{acknowledgements}



\begin{thebibliography}{99}

\bibitem{bc2001}P. Bak and D.R. Chialvo, Phys. Rev. E {\bf{63}} (2001)

\bibitem{bipoo_2001}G-g. Bi and M-m. Poo, Annual Review of Neuroscience {\bf{24}}, 139-166 (2001)
 
\bibitem{blisslomo_1973}T.V.P. Bliss and T. Lomo, J. Physiol. {\bf{232}}, 331-356 (1973)

\bibitem{bp2001}S. Boettcher and A.G. Percus, Phys. Rev. Lett. {\bf{86}}, 5211-5244 (2001)

\bibitem{brooks_1991}R.A. Brooks, Proceedings of the twelfth International Joint Conference on Artificial Intelligence, edited by Morgan Kaufmann 569-595 (1991)

\bibitem{cb1999}D.R. Chialvo and P. Bak, Neuroscience {\bf{90}}, 1137-1148 (1999)

\bibitem{churchland_1992}P.S. Churchland and T.J. Sejnowski, The Computational Brain, MIT Press (1992)

\bibitem{crick_1989}F. Crick, Nature {\bf{337}}, 129-132 (1989)

\bibitem{vphd_2003}F. Emmert-Streib, Aktive Computation in offenen Systemen. Lerndynamiken in biologischen Systemen: Vom Netzwerk zum Organismus. (in german), Dissertation, Universit\"at Bremen (2003)

\bibitem{vhebb3_2003}F. Emmert-Streib, in preperation (2003)

\bibitem{fitzsimonds_1997}R.M. Fitzsimonds, H-j.Song and M-m. Poo, Nature {\bf{388}}, 439-448 (1997)

\bibitem{freymorris_1997}U. Frey and R.G.M. Morris, Nature {\bf{385}}, 533-536 (1997)

\bibitem{h1949}D.O. Hebb, The Organization of Behavior, Wiley, New York (1949)

\bibitem{hertzkrogh_1991}J. Hertz, A. Krogh and R.G. Palmer, Introduction to the theory of neural compuation, Addison-Wesley (1991)

\bibitem{kempter_1999}R. Kempter, W. Gerstner and J.L. van Hemmen, Phys. Rev. E {\bf{59}}, 4498-4514 (1999)

\bibitem{kbs2000}K. Klemm, S. Bornholdt and H.G. Schuster, Phys. Rev. Lett. {\bf{84}}, 3013-3016 (2000)

\bibitem{koch_1999}C. Koch, Biophysics of Computation, Oxford Press (1999)

\bibitem{markramluebke_1997}H. Markram, J. L\"ubke, M. Frotscher and B. Sakmann, B., Science {\bf{275}}, 213-215 (1997)

\bibitem{minskypapert_1969}M. Minsky and S. Papert, Perceptrons, MIT Press (1969)

\bibitem{moore_1920}E.H. Moore, Bull. Amer. Soc., 394--395 (1929)

\bibitem{otmakhova_1998}N.A. Otmakhova and J.E. Lisman, J. Neuroscience  {\bf{18}}, 1270-1279 (1998)

\bibitem{p1955}R. Penrose, Proc. Cambridge Phil. Soc. {\bf{51}}, 405-413 (1955)

\bibitem{rumelharthinton1_1986}D.E. Rumelhart, G.E. Hinton and R.J. Williams, Nature {\bf{323}}, 533-536 (1986)

\bibitem{simon_1996}H.A. Simon, The Science of the Artificial, MIT Press, Cambridge (1996)

\bibitem{varela_1991}F.J. Varela, E. Thompson and E. Rosch, The Embodied Mind, MIT Press (1991)

\bibitem{WattsStro_1998}D.J. Watts and S.H. Strogatz, Nature {\bf{393}}, 440-442 (1998)


\end{thebibliography}

\end{document}